\documentclass[prl,twocolumn,showpacs,preprintnumbers,amsmath,amssymb]{revtex4}

\usepackage{graphicx}

\begin{document}

\title{Magnetic Excitations of Coupled Spin Ladders: a Quantum Monte Carlo Study}

\author{Brian M. Andersen$^1$ and Olav F. Sylju\aa sen$^2$}
\affiliation{$^1$Department of Physics, University of Florida,
Gainesville, Florida 32611-8440, USA\\$^2$NORDITA, Blegdamsvej 17,
DK-2100 Copenhagen \O, Denmark}

\date{\today}

\begin{abstract}

We calculate the magnetic excitation spectrum in the stripe phase of
high-T$_c$ materials. The stripes are modeled as coupled spin-1/2
ladders and the spin dynamics is extracted using Quantum Monte Carlo
(QMC) simulations, which can capture the strong quantum fluctuations
near quantum critical points of coupled spin ladders. We find a
characteristic hourglass magnetic excitation spectrum with
high-energy peaks rotated by 45 degrees compared to the
incommensurate low-energy peaks in good agreement with the
experimental data. The excitations are investigated quantitatively
as a function of interladder coupling, ladder width, and domain
formation with stripe disorder.

\end{abstract}


\maketitle

Neutron scattering experiments are powerful probes of the bulk spin
excitations of magnetic materials, and has been applied extensively
to the cuprates since their discovery\cite{tranqreview}. However,
only recently have observations of universal fluctuations been
reported\cite{tranqLBCO,steven}, providing a crucial new piece of
information for the general understanding of how these materials
evolve from an antiferromagnetic (AF) Mott insulator to a $d$-wave
superconductor as electrons are extracted from the CuO$_2$ planes.

There is increasing evidence that competing interactions in the
cuprate materials result in spatially inhomogeneous electronic spin
and charge ground states in a significant region of the phase
diagram, and it is important to determine whether experimental bulk
probes are in agreement with such inhomogeneous solutions in
general, and one-dimensional stripe configurations in
particular\cite{kivelson}. While it might be tempting to model the
magnetic response by checkerboard order\cite{hanaguri}, it is
well-known that such textures cannot reproduce the
data\cite{andersenprb}. A recent neutron study of
La$_{1.48}$Nd$_{0.4}$Sr$_{0.12}$CuO$_4$ has confirmed that indeed
the magnetic order is collinear and modulated one-dimensionally
consistent with stripes\cite{niels2}.


Recent extensive neutron scattering experiments on several cuprate
materials have been performed to map out the details of the magnetic
fluctuations over a large energy
range\cite{tranqLBCO,steven,stock,hinkov,pailhes}. The compound
La$_{1.875}$Ba$_{0.125}$CuO$_4$ (LBCO) is known to support static
stripe spin and charge order below $T=50$K as indicated by
superlattice Bragg peaks in both sectors\cite{fujita}. The
low-energy magnetic excitation spectrum of this
material\cite{tranqLBCO} (at $T
> T_c$) consists of inwardly dispersing incommensurate spin branches.
In La$_{2-x}$Sr$_x$CuO$_4$ (LSCO) the momentum position of similar
incommensurate modes are known to scale linearly with the doping up
to $\sim \frac{1}{8}$, a property that is naturally explained within
the stripe scenario\cite{yamada,lorenzanaDresden}. For LBCO the
low-energy branches merge into the "resonance point" at $(\pi,\pi)$
at an energy close to 50 meV. At even higher energies, constant
energy cuts reveal that the excitations disperse outward and form a
square or ring feature with four prominent peaks rotated by 45
degrees compared to the low-energy incommensurate modes.
The resulting spin response exhibits a characteristic hourglass
shape, and is remarkably similar to the data obtained in
LSCO\cite{niels} and YBa$_2$Cu$_3$O$_{6+x}$ (YBCO)
\cite{steven,arai,mook,bourgesybco,reznik,stock,pailhes}. This
supports the notion of a universal spin spectrum\cite{batista}
modulo various quantitative differences, particularly at low
energies resulting from e.g. material dependent exchange couplings,
quasiparticle damping effects, and possible existence of a doping
and temperature dependent spin gap.

Theoretically, several groups have used various spin-only models to
compute semiclassically the magnetic excitations in stripe states
with magnetic long-range order (LRO)\cite{batista,kruger,carlson}.
It has, however, been pointed out that strong quantum fluctuations
are important for obtaining the correct high-energy response
originating from the quasi-one-dimensional spin excitations of the
ladder\cite{tranqLBCO,vojta,uhrig,seibold}, even though, in certain
limits, the magnetic fluctuations can closely resemble the
semiclassical results\cite{yao}. The importance of the charges in
describing the spin response remains controversial, but is expected
to affect the details at low energy as mentioned above. The charge
degrees of freedom have been included within phenomenological
models\cite{vojtasachdevjpcs}, and within time-dependent Gutzwiller
approximation of the Hubbard model\cite{seibold}. The coexistence
phase of static stripes and $d$-wave superconductivity was studied
in Ref. \onlinecite{andersen} using a mean-field+RPA approximation.

In this paper, motivated by the discovery of universal spin
fluctuations, we calculate the dynamical structure factor
$S({\mathbf{q}},\omega)$ with a powerful QMC method able to treat
both the ordered and quantum disordered regimes. We discuss effects
of interladder coupling, ladder width, and disorder. As opposed to
previous microscopic spin-only models, we do not rely on
semiclassical linearized spin-wave\cite{batista,kruger,carlson,yao},
or one-triplon\cite{uhrig} approximations.

The spin dynamics is qualitatively different in isolated even and
odd leg spin-1/2 ladders\cite{dagottorice}; whereas even leg ladders
display a spin gap and short-range spin-spin correlations, the odd
leg ladders are qualitatively similar to the spin-1/2 chain with
gapless quasi-long-range correlations. For coupled ladders, the odd
leg ladders support an ordered state for any interladder exchange
coupling $J_b$, whereas coupled even leg ladders exhibit a quantum
critical point (QCP) at some finite value of $J_b$\cite{kim}. For
the cuprates, at a doping near $x = \frac{1}{8}$ the charge is
modulated with a period close to four lattice constants, half the
period of the associated spin modulations. Hence, the associated
stripe phase may consist of coupled 2-, 3-, or 4-leg spin ladders
containing 2, 1, and 0 spin-empty sites, respectively. In the
following, the spins on the Cu atoms in the CuO$_2$ planes are
modeled by the Heisenberg Hamiltonian defined on a square lattice
\begin{equation}
\mathcal{H}=\sum_{\langle ij \rangle} J_{ij} {\mathbf{S}}_i \cdot
{\mathbf{S}}_j\label{heisenberg},
\end{equation}
where ${\mathbf{S}}_i$ is the spin-1/2 operator at site $i$, and
$J_{ij}$ denotes the exchange coupling between sites $i$ and $j$.
Uni-directional stripes parallel to the $y$-axis are modeled by
couplings $J_b$ across the hole-rich quasi-one-dimensional regions
and $J_a$ elsewhere\cite{riera}. The Hamiltonian (\ref{heisenberg})
and the assumed stripy configurations of exchange couplings is the
resulting effective spin-only model after integrating out the charge
carriers. Whereas $J_a > 0$, the sign of $J_b$ depends on the period
of the stripes: $J_b > 0$ for site-centered stripes and $J_b < 0$
for bond-centered stripes\cite{yao}. The choices accommodate the
anti-phase property of the stripes believed to originate from a
lowering of the kinetic energy of the carriers. Below, we study
$L\times L$ lattices with periodic boundary conditions, and use
tetragonal units with wave vectors $q_x$, $q_y$ parallel to the Cu-O
bonds, and energies are given in units of $J_a$. We typically set
$L=64$ allowing us to study momentum resolved correlations functions
within QMC.

The QMC simulations were performed using the Stochastic Series
Expansion (SSE) method\cite{SSE} with directed-loop
updates\cite{SS}. In order to extract imaginary-time correlation
functions the individual vertices in an SSE configuration were each
assigned a random imaginary-time location in such a way that the
assignment did not alter the imaginary-time ordering sequence of
vertices\cite{INTERACTION}. The $\langle S^z_i(\tau) S^z_j(0)
\rangle$ correlation function was read out using a mesh with 200
grid points along the imaginary-time axis and Fast Fourier
transforms were used to obtain the momentum space points. The
imaginary-time data was finally continued to real-frequency using
the MaxEnt method\cite{MaxEnt} with a flat prior.

\begin{figure}[t]
\begin{minipage}{.49\columnwidth}
\includegraphics[clip=true,width=.95\columnwidth,height=4.0cm]{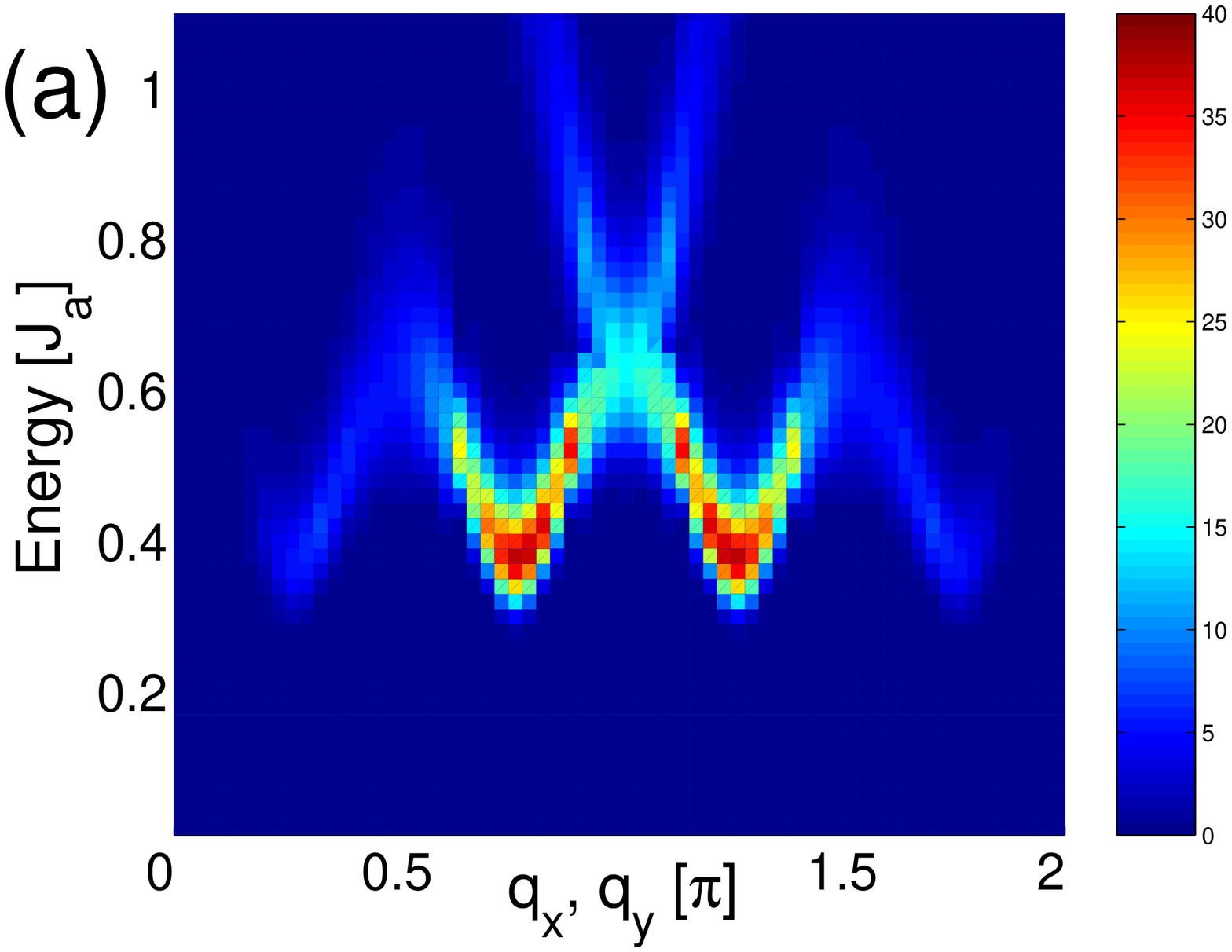}
\end{minipage}
\begin{minipage}{.49\columnwidth}
\includegraphics[clip=true,width=.95\columnwidth,height=4.0cm]{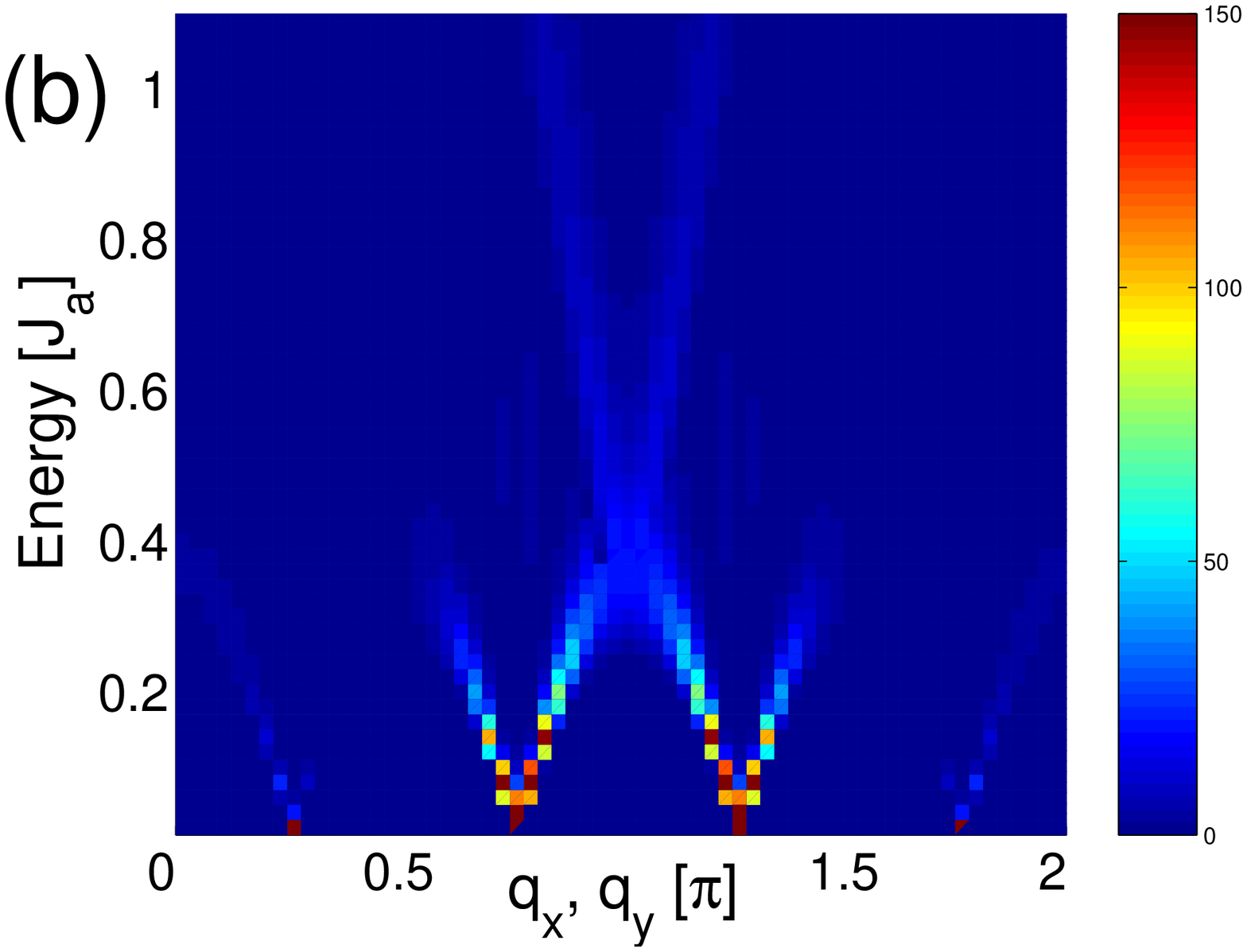}
\end{minipage}\\
\begin{minipage}{.49\columnwidth}
\includegraphics[clip=true,width=.95\columnwidth,height=4.0cm]{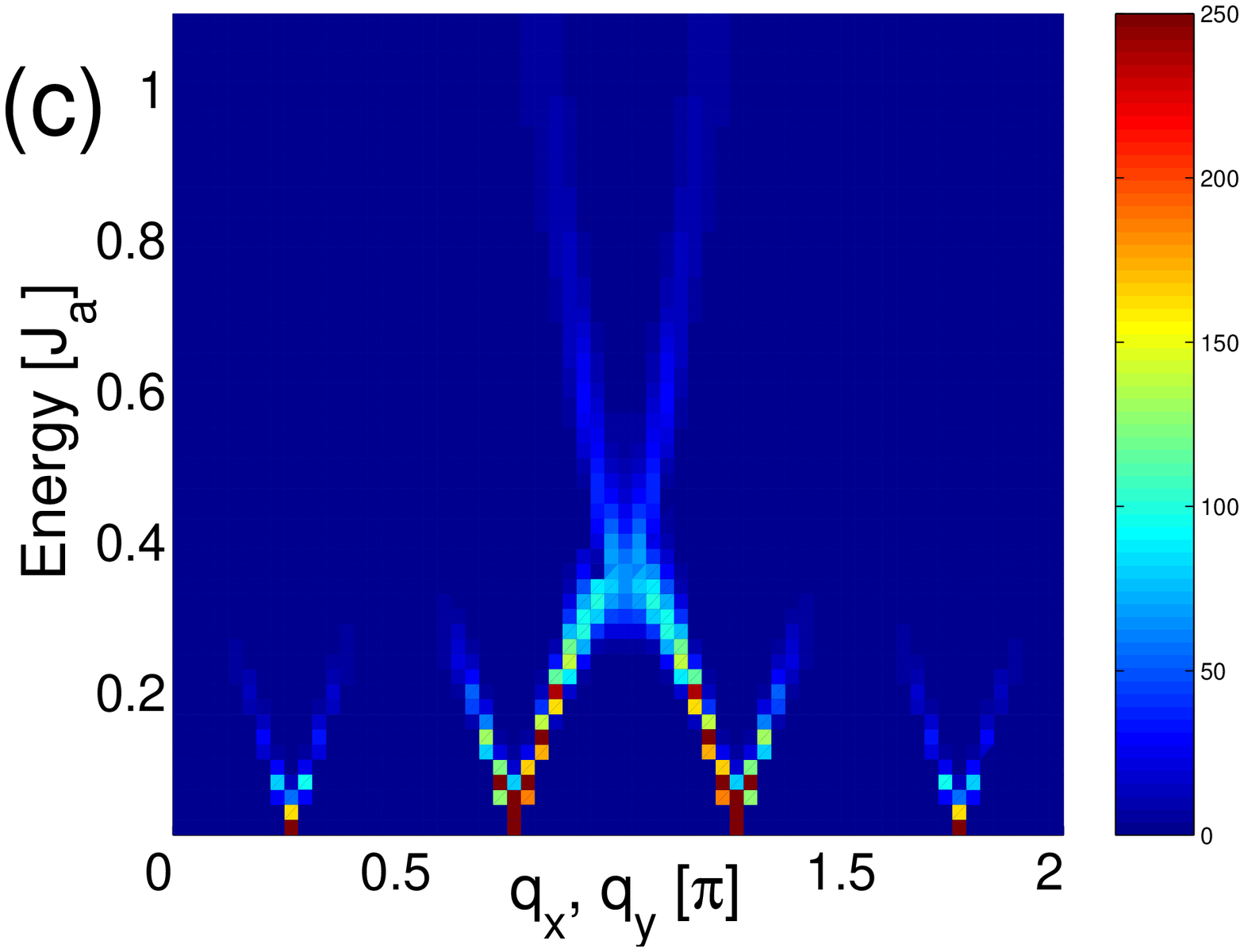}
\end{minipage}
\begin{minipage}{.49\columnwidth}
\includegraphics[clip=true,width=.95\columnwidth,height=4.0cm]{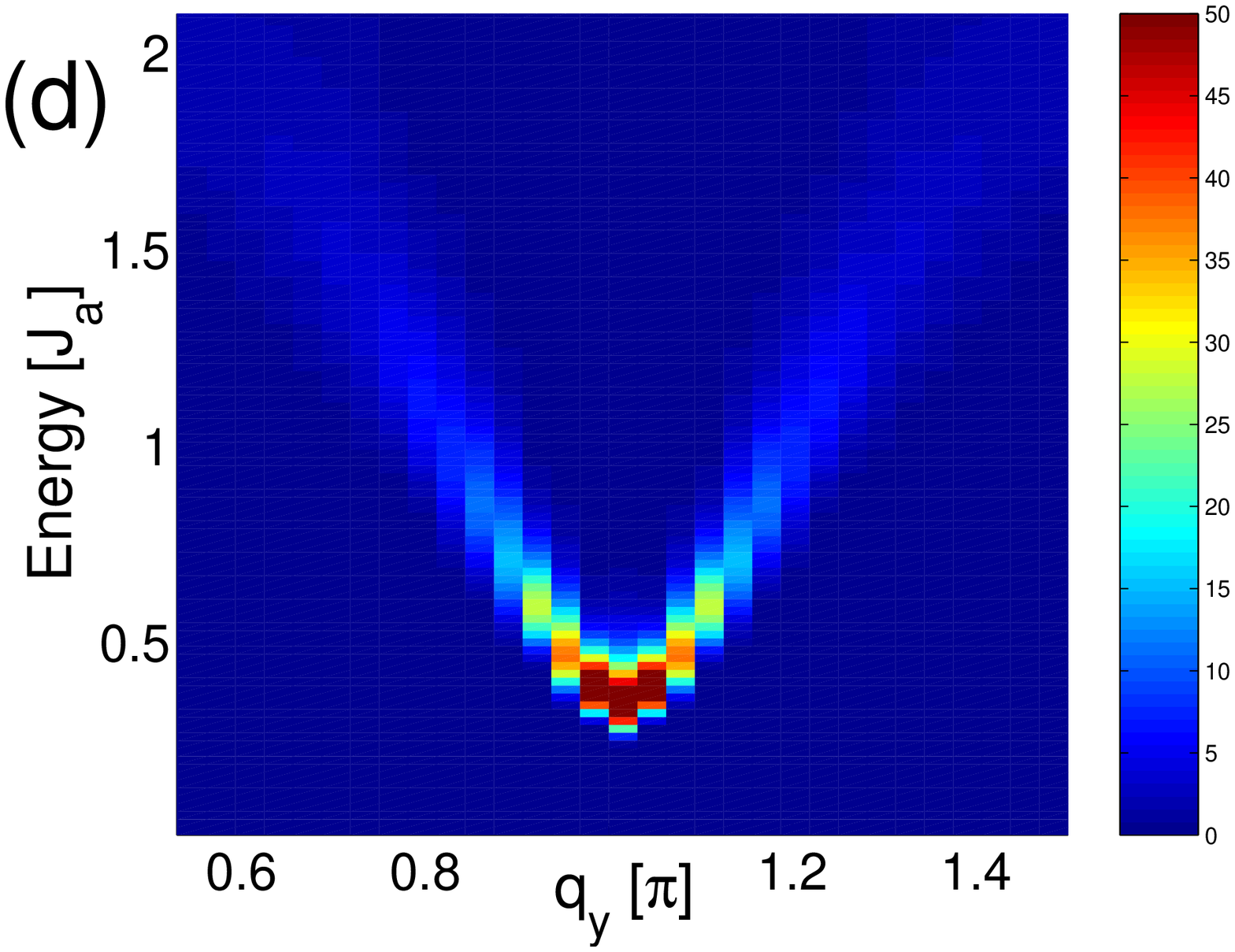}
\end{minipage}
\caption{(Color online) (a-c) $S(q_x,\pi,\omega)$ vs $q_x$
superimposed with $S(\pi,q_y,\omega)$ vs $q_y$ showing the
characteristic X-shaped hourglass spin excitation spectrum for
coupled 2- (a), 3- (b), and 4-leg ladders (c) with $|J_b|=0.1$ and
$J_a\beta=80$. (d) $S(\pi,q_y,\omega)$ vs $q_y$ from (c) showing the
high-energy quasi 1D dispersion along the 4-leg spin ladder (note
different scale). In (b,c) the amplitude of peaks at $(1\pm
\frac{3}{4})\pi$ are reduced by an order of magnitude compared to
the main peaks at $\left(1\pm \frac{1}{4}\right)\pi$.}
\label{fig:hourglass}
\end{figure}
\begin{figure}[t]
\begin{minipage}{.49\columnwidth}
\includegraphics[clip=true,width=.95\columnwidth]{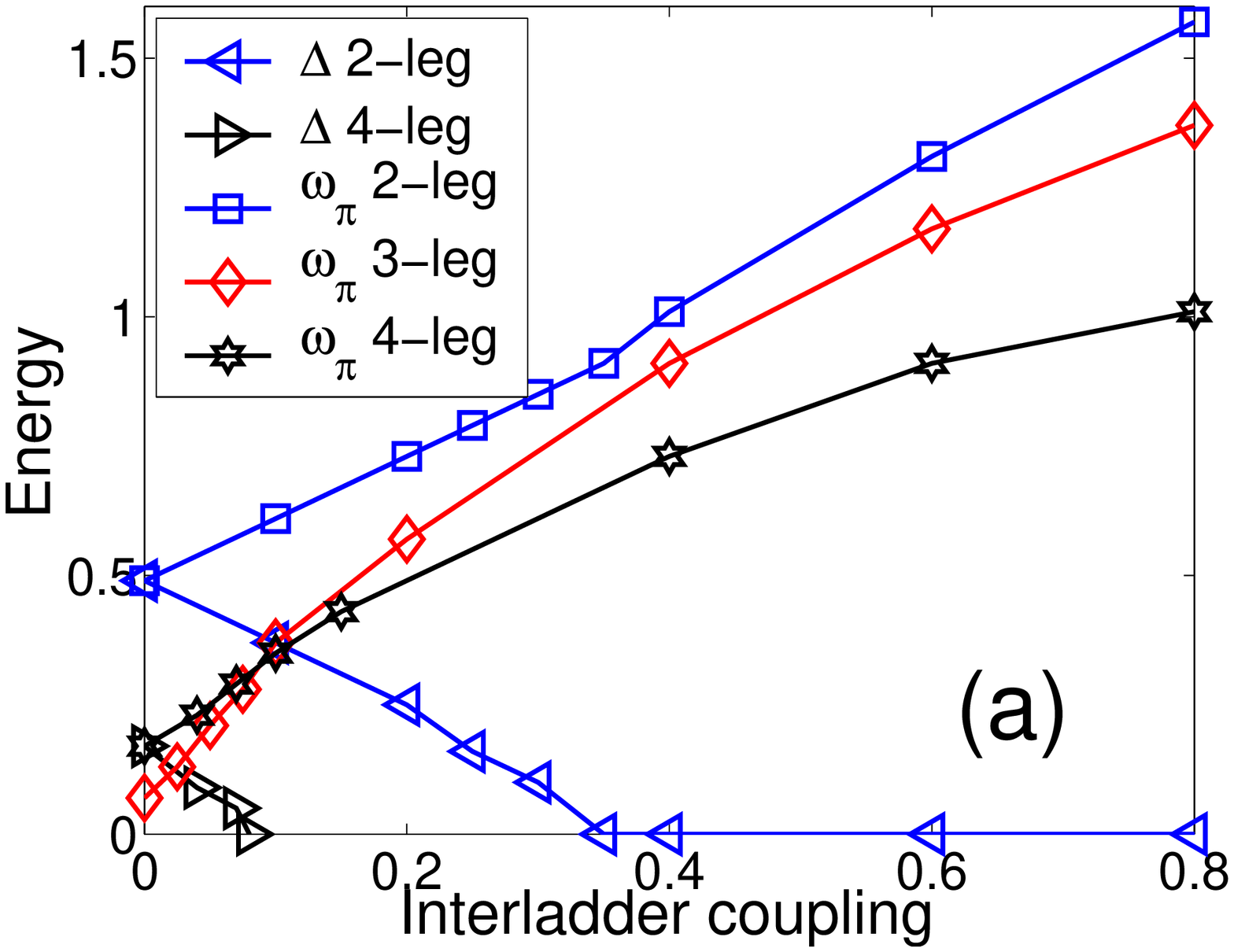}
\end{minipage}
\begin{minipage}{.49\columnwidth}
\includegraphics[clip=true,width=.95\columnwidth]{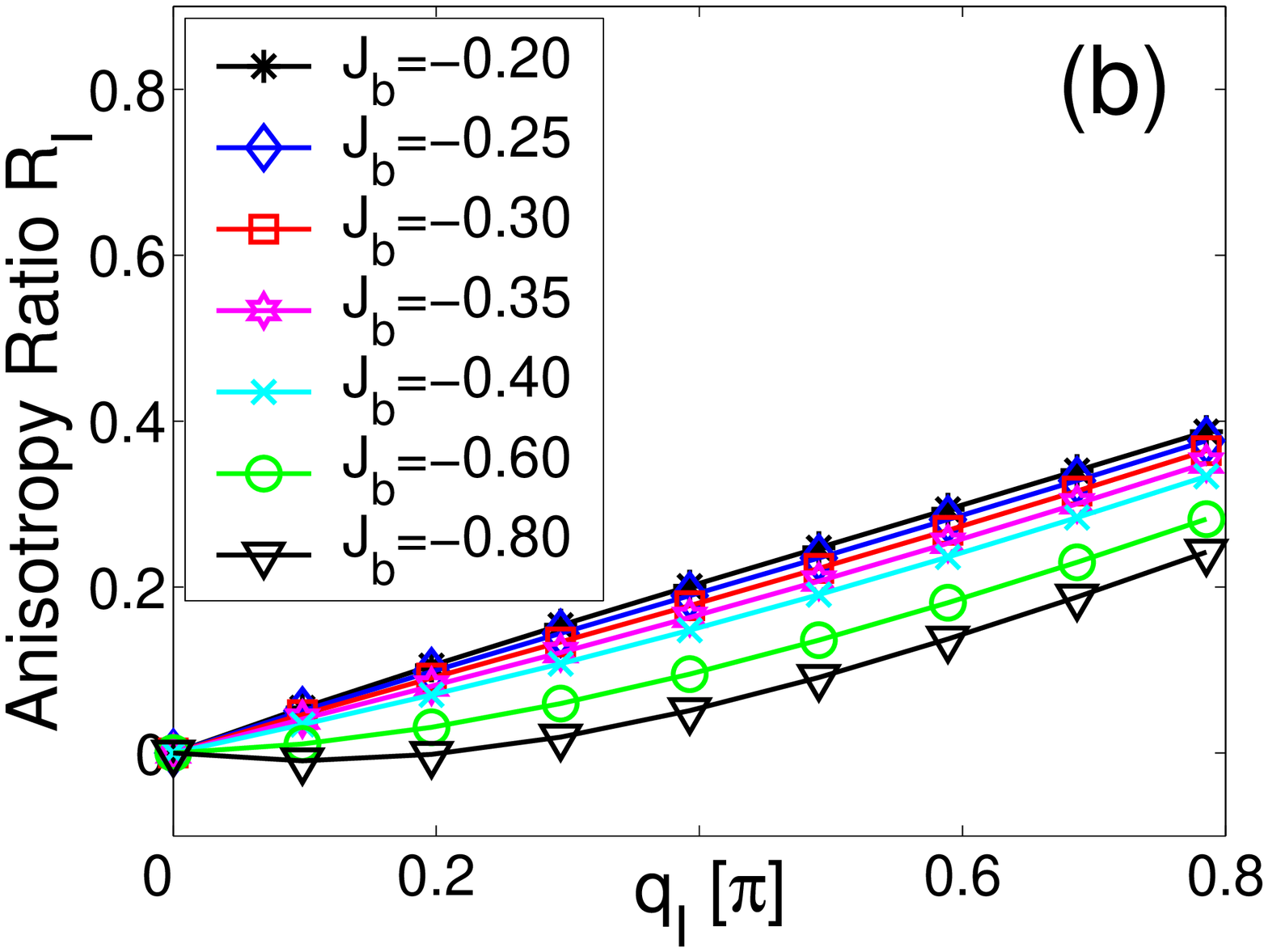}
\end{minipage}\\
\begin{minipage}{.49\columnwidth}
\includegraphics[clip=true,width=.95\columnwidth]{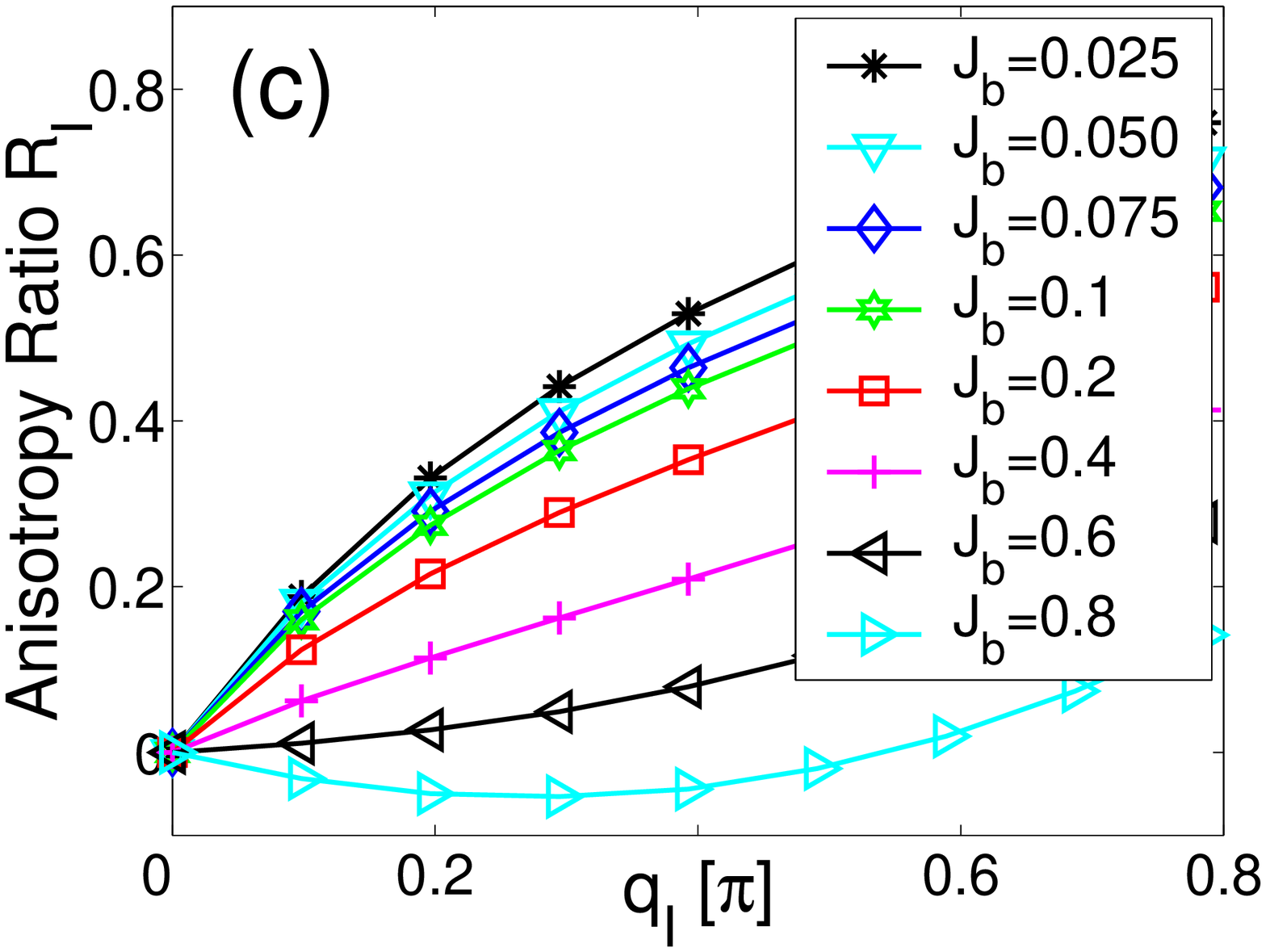}
\end{minipage}
\begin{minipage}{.49\columnwidth}
\includegraphics[clip=true,width=.95\columnwidth]{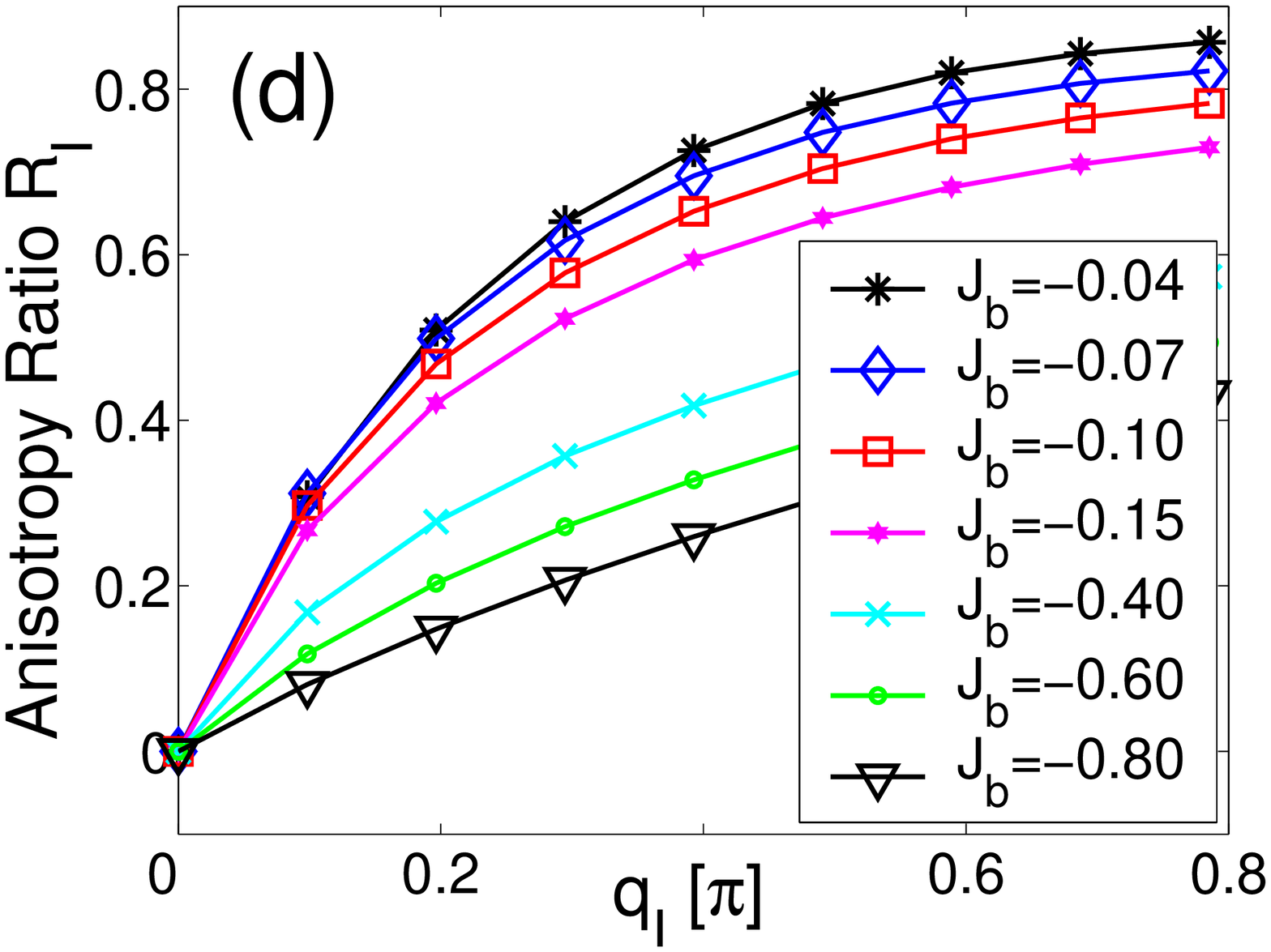}
\end{minipage}
\caption{(Color online) (a) Resonance $\omega_\pi$ and mode gap
$\Delta$ at $3\pi/4$ vs interladder coupling $J_b$. (b-d) The
anisotropy ratio $R_I$ vs $q_I=|q_B-q_x|$ for coupled 2- (b), 3-
(c), and 4-leg ladders (d). All results are obtained with
$J_a\beta=40$ and $L=64$.} \label{fig:resonance}
\end{figure}

Fig.~\ref{fig:hourglass} shows superimposed results for
$S({\mathbf{q}},\omega)$ as a function of energy $\omega$ and
momenta along cuts $(q_x,\pi)$ and $(\pi,q_y)$ for coupled 2- (a),
3- (b) and 4-leg ladders (c), all with $|J_b|=0.1$ for easy
comparison. For this value of $J_b$ the coupled 2-leg ladders are in
the quantum disordered regime and displays a gap as seen in
\ref{fig:hourglass}(a), whereas magnetic LRO and Goldstone modes
exist for the coupled 3- and 4-leg ladders.
Fig.~\ref{fig:hourglass}(a-c) clearly show an hourglass spin
spectrum consisting of anisotropic low-energy spin modes and a
dispersive high-energy branch characteristic of
quasi-one-dimensional quantum paramagnets, as shown separately in
Fig.~\ref{fig:hourglass}(d). Important features in
Fig.~\ref{fig:hourglass} are the energy $\omega_\pi$ of the saddle
point in the dispersion at $(\pi,\pi)$ (resonance point), and the
minimum gap $\Delta$ at $\left(1\pm \frac{1}{4}\right)\pi$.
Fig.~\ref{fig:resonance}(a) shows the resonance energy $\omega_\pi$
for the 2-, 3-, and 4-leg ladders as a function of the interladder
coupling $|J_b|$. As seen, $\omega_\pi$ increases with $|J_b|$, and
the even leg ladders display a gap at $J_b=0$ as expected. Note that
for the 3-leg ladder $\omega_\pi$ does not vanish identically at
$J_b=0$ due to the finite temperature. This agrees with
$\sigma$-model calculations which give $\omega_\pi=\pi T + ...$,
where ellipses denote higher order logarithmic
corrections\cite{olav}.

The intensity anisotropy ratio $R_I$ between the inner (toward
$(\pi,\pi)$) and outer (away from $(\pi,\pi)$) branches of the
low-energy modes is another important quantity since the outer
branches are not yet observable in experiments. We define an
anisotropy ratio by
$R_I=\left[S(q_B+q_I,\pi)-S(q_B-q_I,\pi)\right]/S(q_B+q_I,\pi)$
where $S({\mathbf{q}})$ is the energy integrated structure factor,
and $q_I=|q_B-q_x|$ with $q_B=3\pi/4$ one of the Bragg points.
Fig.~\ref{fig:resonance}(b-d) show $R_I$ vs $q_I$ for the coupled
2-, 3-, and 4-leg ladders for various values of $J_b$. Clearly $R_I$
decreases with increasing interladder coupling\cite{yao}, and for
the 2-leg ladders $R_I$ remains small ($\lesssim 15 \%$) in the
vicinity of the Bragg point for all $J_b$. In fact, based only on
$R_I$ one would expect the 4-leg ladders to best model the
experiments\cite{stock,hinkov,pailhes,fujita,niels,arai,bourgesybco,mook,reznik}.
For large enough $J_b$, $R_I$ eventually becomes negative and the
outer branches dominate the intensity.
We have also studied coupled 6- and 8-leg ladders, and found
excitations qualitatively similar to Fig.~\ref{fig:hourglass}(b-c)
but with even larger anisotropy ratios between the inner and outer
low-energy spin branches.

\begin{figure}[t]
\begin{center}
\includegraphics[width=0.99\columnwidth,height=9.5cm]{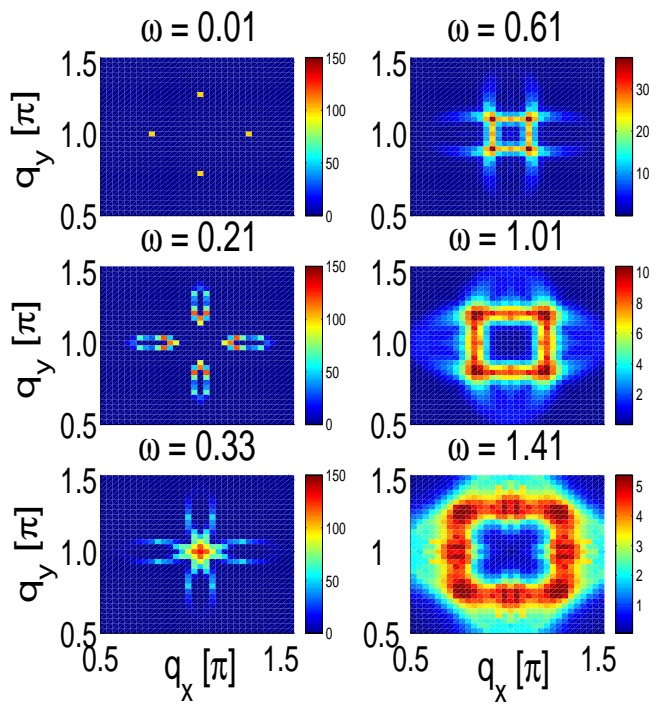}
\end{center}
\caption{(Color online) Constant energy cuts of
$S({\mathbf{q}},\omega)$ for coupled 4-leg ladders; $J_b=-0.1$,
$L=64$, and $J_a \beta=80$. Similar results can be generated for
coupled 2- and 3-leg ladders but with reduced anisotropy ratios
$R_I$ (Fig.~\ref{fig:resonance}).}\label{fig:cuts}
\end{figure}

Fig.~\ref{fig:cuts} shows constant energy cuts of
$S({\mathbf{q},\omega})$ for the coupled 4-leg ladders with the same
parameters as in Fig.~\ref{fig:hourglass}(c-d). Similar to
experiments, with increasing energy the incommensurate low-energy
peaks merge at the resonance point around $\sim 0.3$ and the
high-energy response consists of a diamond structure dominated by
four peaks rotated 45 degrees compared to the ones at low-energy.
\begin{figure}[b]
\begin{center}
\includegraphics[width=0.95\columnwidth,height=5.0cm]{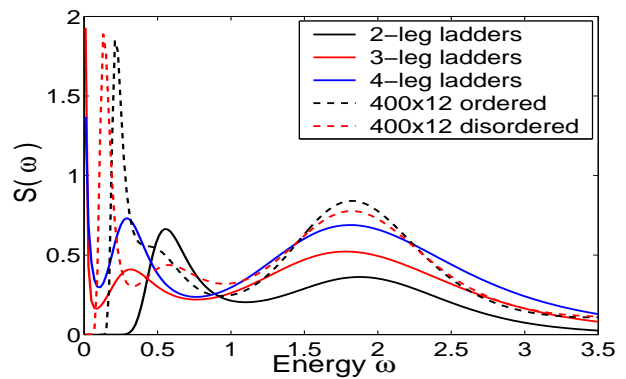}
\end{center}
\caption{(Color online) $S(\omega)$ for coupled 2- (black) 3- (red)
and 4-leg (blue) ladders with parameters similar to Fig.
\ref{fig:hourglass}. Dashed lines show the results for $400\times
12$ sites with ordered (black) and disordered (red) 4-leg ladders
($J_b=-0.1$). }\label{fig:integrated}
\end{figure}
In Fig.~\ref{fig:integrated} we show the momentum integrated
structure factor $S(\omega)$ for the same parameters used in Fig.
\ref{fig:hourglass}. Besides the $\omega=0$ Bragg peak (for the
ordered cases), $S(\omega)$ is dominated by two characteristic peaks
corresponding to the saddle point in the dispersion at the resonance
point, and a peak at the upper band edge of the dispersive branch
along the ladders ($\omega\sim 2J_a$).

Important open questions relates to the effects of disorder and
stripe domain formation on the magnetic excitations, and the origin
of the spin gap. In LSCO magnetic LRO is observed at low doping
$x\lesssim 0.14$\cite{tranq1995,yamada}, whereas at higher doping a
spin gap phase, which may be related to superconductivity, opens up.
YBCO and BSCCO exhibit a large spin gap, i.e. the stripes remain
fluctuating at all doping levels possibly related to a smaller
interladder coupling $J_b$ leaving these materials in the quantum
disordered regime. Within a phenomenological approach Vojta {\sl et
al.}\cite{vvk} recently studied the spin susceptibility in the
presence of slow dynamically fluctuating charge stripes. As a
further complexity there is also evidence from muon spin relaxation
experiments that LSCO and YBCO exhibit a spin-glass phase with local
'static' short-range order.

\begin{figure}[b]
\begin{minipage}{.49\columnwidth}
\includegraphics[clip=true,width=.95\columnwidth,height=4.2cm]{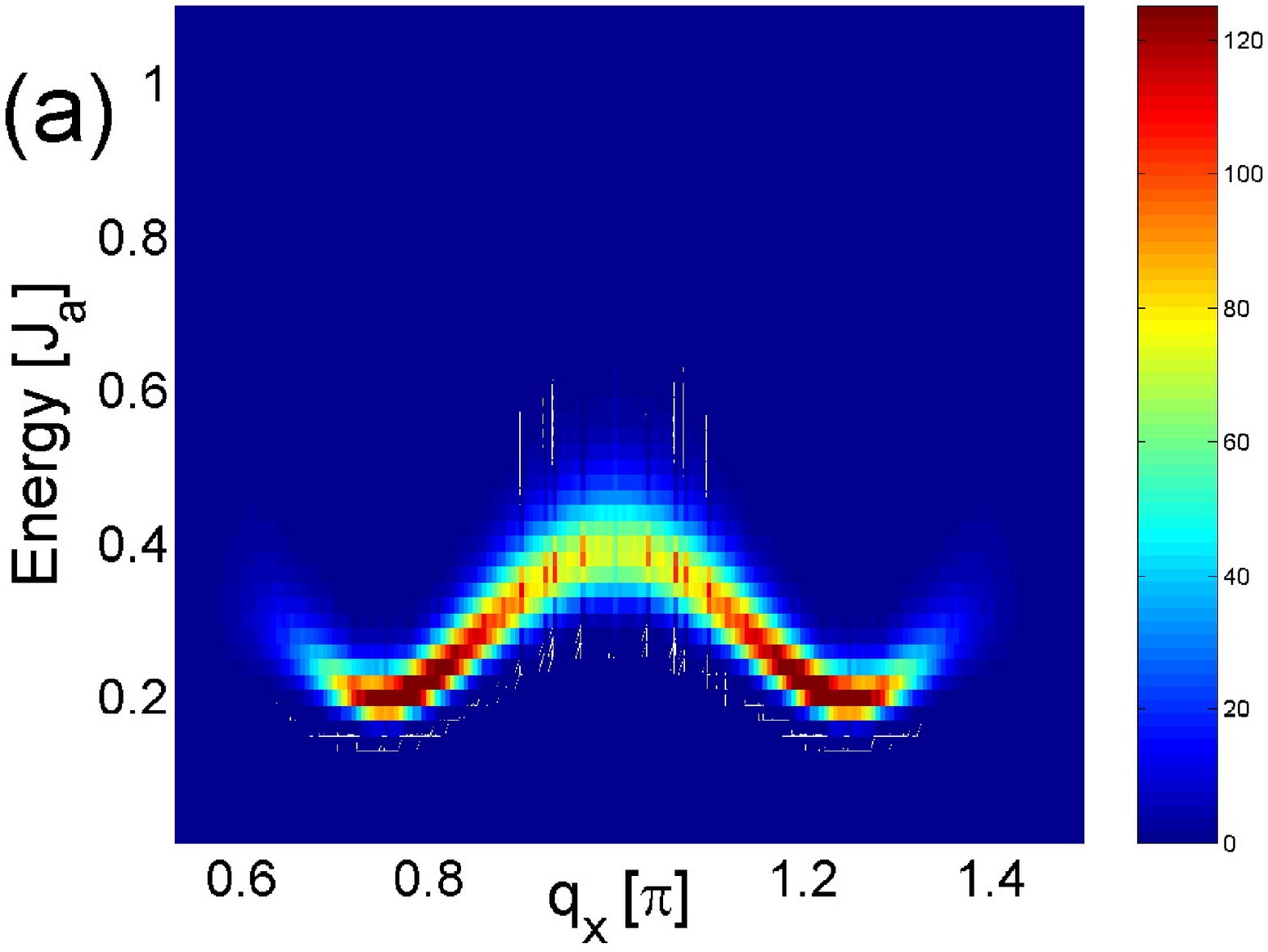}
\end{minipage}
\begin{minipage}{.49\columnwidth}
\includegraphics[clip=true,width=.95\columnwidth,height=4.2cm]{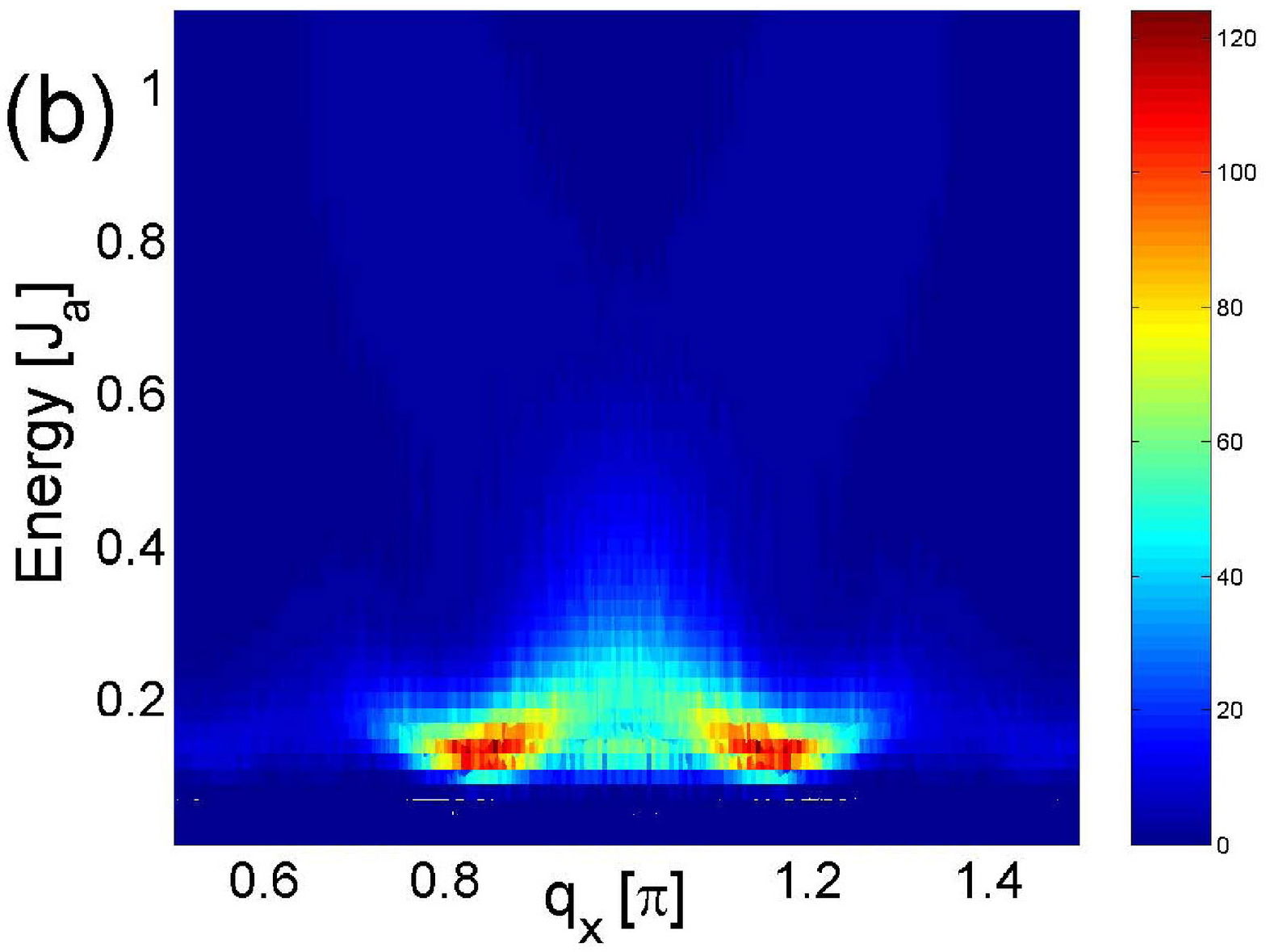}
\end{minipage}
\caption{(Color online) "Boomerang" excitation spectrum for a
$400\times 12$ system of ordered (a) and disordered (b) stripes.}
\label{fig:disordered400x12}
\end{figure}

In the rest of this paper, we wish to study a simple disorder
scenario which simulates both random stripe positions and finite
sized stripe domains. Domain formation was an assumption when
symmetrizing the plots in Figs.~\ref{fig:hourglass} and
\ref{fig:cuts}, and should be a natural consequence of random
quenched disorder\cite{robinson}. In the following we simulate this
kind of disorder by studying systems of size $400\times L$ where $L$
is an assumed average domain size, and the stripe positions are
disordered with a flat distribution in the interval [1,7] (mean 4).
Clearly this approach is crude, but has been successfully applied to
explain the quasiparticle spectral weight as detected by
photoemission spectroscopy\cite{granath}. In general, we find that
this kind of disorder is very detrimental to the hour-glass
excitations {\sl unless} $L$ is small enough to induce a spin gap.
For example, for $400\times 12$ systems with parameters similar to
Fig.~\ref{fig:cuts}, we show in Fig.~\ref{fig:disordered400x12} the
result for ordered (a) and disordered (b) stripes. The disordered
case is averaged over 4 different configurations which is enough for
these large systems. The corresponding $S(\omega)$ are shown in
Fig.~\ref{fig:integrated}.
Fig.~\ref{fig:disordered400x12} demonstrates a case of
domain-induced spin gap which is a property of the individual $4
\times 12$ clusters that (when coupled) constitute the system (a gap
is negligible in the $64\times 64$ system (Fig.~\ref{fig:cuts}), and
in $400\times 12$ systems with $J_b\sim J_a$ (not shown)).
Interestingly, as seen from Fig.~\ref{fig:disordered400x12}(b), the
stripe disordering not only smears the excitation spectrum resulting
in diffuse 'legs of scattering', but also strongly reduces the spin
gap. In this case the resulting pile-up of weight right above the
spin gap is clearly unrelated to (but may be enhanced
by\cite{andersen}) superconductivity. We expect this scenario to be
most relevant for LSCO\cite{niels}, whereas in YBCO, where the
intensity is peaked near $(\pi,\pi)$, fluctuating charge
stripes\cite{vvk} or superconductivity\cite{andersen} may be
important. For this material, however, it remains controversial
whether homogeneous models are better starting points, at least for
the optimally-to overdoped regime\cite{manske,eremin}.

In summary, we have calculated the spin excitations in the stripe
phase within a Heisenberg model of coupled spin ladders using QMC,
and investigated the effects of interladder coupling, ladder width
and simple types of disorder. In agreement with experiments, we find
low-energy spin-wave like excitations with intensity strongly
dominated by the inner branches for low interladder coupling. These
merge at the $(\pi,\pi)$ point and rotate into the quantum
excitations characteristic of the quasi-one-dimensional spin ladders
at higher energies.

\end{document}